\documentclass{article}

\usepackage[english]{babel}

\usepackage[letterpaper,top=2cm,bottom=2cm,left=3cm,right=3cm,marginparwidth=1.75cm]{geometry}

\usepackage{amsmath}
\usepackage{caption}
\usepackage{graphicx}
\usepackage[utf8]{inputenc}
\usepackage{booktabs}
\usepackage{multirow}
\usepackage{longtable}
\usepackage{array}
\usepackage{geometry}
\usepackage[colorlinks=true, allcolors=blue]{hyperref}
\usepackage{authblk}

\title{Open-source segmentation and biometry dataset using spectrally-multiplexed whole-eye optical coherence tomography}
\author{Ruobing Qian*, Catherine Fromm*, Pushkar Anand, Kyle Johnson, Zach Willms \\
Yimin Ding, Weihan Zhang, Ali Behrooz, Mohamed El-Haddad}

\affil{Reality Labs, Meta}
\begin{document}
\maketitle

\begin{abstract}

Whole-eye optical coherence tomography (WEOCT) has emerged as a transformative imaging modality capable of simultaneously capturing the anterior and posterior segments of the human eye. WEOCT enables comprehensive ocular biometry, which is critical for a wide range of clinical and research applications—from intraocular lens power calculation, myopia progression monitoring, and refractive surgery planning to the precise measurement of the visual and optical axes and the generation of personalized eye models for eye tracking in virtual, augmented and mixed reality (VR/AR/MR). However, existing WEOCT systems often face trade-offs between signal-to-noise ratio (SNR), imaging speed, and the ability to capture dynamic processes without motion artifacts. To address these limitations, we present a novel spectrally-multiplexed WEOCT system that utilizes two synchronized 200 kHz swept sources at 1310 nm and 1060 nm. Coupled with an automated  end-to-end processing pipeline involving deep learning-based surface segmentation, 3D distortion correction, surface fitting and ray-tracing refraction correction, our system enables anatomically accurate 3D reconstruction of the segmented ocular layers. Through a 300+ participant user study and comprehensive phantom studies, we demonstrate that our system can provide simultaneous accurate  measurements of cornea topography and 3D pupil center. While labeled retinal OCT data is abundantly available in open-source repositories, labeled B-scan or volumetric anterior segment data remains significantly limited. Consequently, research groups working in related domains must often acquire their own data using custom imaging systems. To help bridge this gap,  we  are releasing as open-source a comprehensive dataset comprising 6,621 processed volumes from 276 unique participants with corresponding segmentation and calibrated 3D anterior point clouds\footnote{Dataset available at \url{https://github.com/facebookresearch/rl_whole_eye_oct}, see Section \ref{sec:dataset} for details}. To our knowledge, this is the largest publicly available whole-eye OCT datasets along with calibrated 3D anterior point clouds. 

\end{abstract}

\section{Introduction}
Optical coherence tomography (OCT) has become a standard tool in ophthalmic care, vision research, and ocular disease diagnosis, providing high-resolution, cross-sectional imaging of ocular microstructure with micrometer-scale resolution \cite{huang1991optical}. Broadly, OCT systems for eye imaging are divided into two categories based on the anatomical region they target: anterior segment OCT, which is designed to image the cornea, iris, crystalline lens and anterior chamber; and retinal OCT, which images the posterior segment including the macula, optic nerve head, and retinal layers. These two system types differ fundamentally in their optical design — anterior segment systems focus light directly onto the cornea, whereas retinal systems must pivot the beam through the pupil in a fan-shaped configuration and compensate for the refractive power of the cornea and crystalline lens in order to focus on the retina. As a consequence of these distinct optical architectures, a conventional OCT system is designed for one purpose only: it can image either the anterior segment or the retina, but not both, unless the imaging optics and OCT engine (e.g. reference arm setting) are physically reconfigured between acquisitions. 

Whole-eye optical coherence tomography (WEOCT) has emerged to bridge this divide, offering a uniquely powerful imaging modality capable of capturing both the anterior and posterior segments simultaneously. The ability to acquire comprehensive, global ocular biometry in a single coordinate system has profound implications across ophthalmology and vision science \cite{kuo2019advances}. In cataract and refractive surgery, accurate intraocular lens (IOL) power calculation relies heavily on precise measurements of axial length, anterior chamber depth, and corneal power \cite{olsen2007calculation}. While neither anterior nor retinal OCT alone can provide axial length, WEOCT systems have demonstrated the ability to capture these metrics comprehensively \cite{grulkowski2012retinal}. Furthermore, WEOCT enables the direct, in vivo measurement of the primary ocular axes—the visual axis (VA), optical axis (OA), and pupillary axis (PA)—as well as the critical angles between them, such as angle alpha and angle kappa \cite{kim2018full}. Accurate measurement of these angles is essential for preventing IOL decentration and minimizing optical aberrations following refractive surgery \cite{park2012measurement}. Beyond surgical planning, WEOCT is highly valuable for disease monitoring and physiological studies. In the management of myopia, axial length elongation is the primary biomarker for disease progression \cite{hou2018axial}. In glaucoma management, WEOCT offers the potential to concurrently assess anterior chamber angle configuration and retinal nerve fiber layer thickness, providing a holistic view of the disease \cite{kuo2019advances}.

While the clinical applications of WEOCT are vast, the modality is also proving to be a critical enabling technology for advanced human-computer interaction, specifically in the development of robust eye tracking (ET) systems for virtual reality (VR) and augmented reality (AR) devices. As spatial computing evolves, gaze-based input has been established as a primary interaction mode. However, the performance of current ET systems is fundamentally limited by the accuracy of the underlying eye biometry models. A personalized eye model containing accurate three-dimensional (3D) anatomical information is required for robust ET performance, particularly for the "tail end" of the user population who experience the highest error rates \cite{qian2025ground}. Because the human eye views objects along the visual axis—which is typically tilted by approximately 5° relative to the optical axis—accurate gaze estimation requires precise knowledge of this angular offset relationship \cite{simpson2026relating}. WEOCT is uniquely suited to provide this ground-truth biometry, as it can simultaneously measure cornea topography, the 3D pupil center, and the fovea location to directly determine the visual and optical axes.

Despite its immense potential, developing a WEOCT system that can capture the entire eye simultaneously without compromising image quality remains challenging. Existing WEOCT technologies generally fall into two categories: sequential imaging and simultaneous imaging. Sequential systems often employ tunable lenses or mechanical switches to adjust focus between the anterior segment and the retina \cite{nankivil2015handheld,ruggeri2012imaging,grulkowski2018swept,kim2016high,hu2024megahertz,urizar2023optical}. However, these systems are fundamentally limited by their inability to measure the visual and optical axes at the exact same instant, which is essential for capturing true dynamic relationships without motion artifacts. Conversely, simultaneous imaging approaches have been developed using techniques such as single-channel long imaging range with vertical-cavity surface-emitting lasers (VCSELs) \cite{grulkowski2012retinal}, dual-depth polarization encoding \cite{kim2018full,jeong2012spectral,dhalla2012simultaneous,mcnabb2018wide}, spectral-domain dual-channel using a beam splitter\cite{dai2012optical}, and spectral-domain multiplexing at 840 nm and 1050 nm \cite{Fan:15}. While these methods allow for concurrent imaging, they often suffer from limited signal-to-noise ratio (SNR) due to maximum-permissible exposure (MPE) limits or inherent sample arm transmission losses, and their relatively slow imaging speeds increase susceptibility to motion artifacts.

To overcome these limitations and optimize both SNR and imaging speed, we have developed a novel spectrally-multiplexed swept-source WEOCT system. Our approach utilizes two distinct swept sources operating at 1310 nm and 1060 nm to simultaneously image the anterior segment and the retina at 200 kHz. The choice of wavelengths is motivated by well-established optical properties of ocular tissue: 1310 nm offers a higher maximum permissible exposure  limit compared to shorter wavelengths, enabling greater illumination power and thus higher SNR for anterior segment imaging, while 1060 nm is well known for achieving high SNR retinal imaging with minimal signal attenuation due to water absorption. In this paper, we detail the optical and system design of this WEOCT engine and describe the automated processing pipeline developed for 3D biometry extraction. We present representative imaging results demonstrating the system's capability to accurately measure cornea topography and 3D pupil center.

Furthermore, to support and accelerate ongoing research in ocular biometry and eye tracking, we provide an open-source comprehensive dataset of annotated, quality-assured whole-eye OCT data alongside this publication.To the best of our knowledge, this constitutes the first publicly available open-source whole-eye OCT dataset, and the first OCT dataset of any kind to provide processed, distortion- and refraction-corrected 3D point clouds of the anterior eye with corresponding anatomical labels.

\section{Methods}

\subsection{OCT system design \& performance} 
\label{sec:oct_engine}
\begin{figure}[h]
    \centering  
    \includegraphics[scale =0.9]{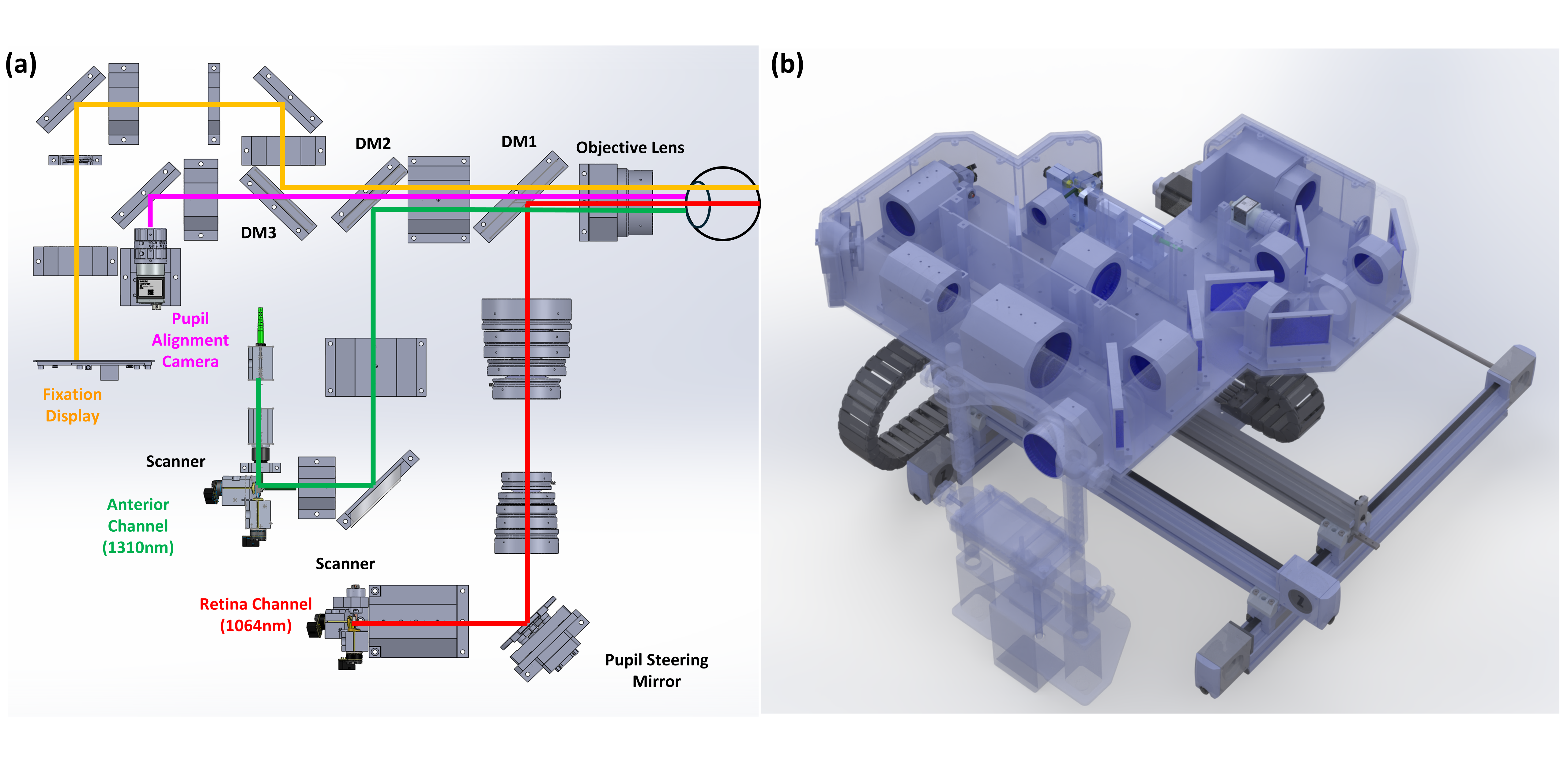}
    \caption{\textbf{Spectrally-multiplexed whole-eye OCT system design (a)layout and (b)CAD model 3D rendering.} Red path: 1060nm retinal OCT channel, green path: 1310nm OCT anterior channel, pink path: pupil alignment camera channel, yellow path: fixation display path, DM: custom dichroic mirror  }
    \label{fig:system}
\end{figure}

Our spectrally-multiplexed whole-eye OCT system (Fig.\ref{fig:system}) employs two independent 200 kHz swept-source lasers (Thorlabs, Newton, NJ) and two high-speed digitizers (ATS9373, Alazar Technologies Inc., Pointe-Claire, Quebec). The anterior channel uses swept source centered at 1310 nm with a 60 nm bandwidth. With dual-edge sampling, the digitizer captures 6,656 k-clock samples per sweep, yielding a maximum imaging depth of approximately 50 mm in air—sufficient to image the entire anterior segment from the eyelashes through the cornea to the posterior surface of the crystalline lens. The retina channel uses a swept source centered at 1064 nm with a 102 nm bandwidth, capturing 4,096 k-clock samples per sweep after dual-edge sampling, corresponding to a maximum imaging depth of 12.2 mm in air, which ensures retinal coverage across subjects with varying ocular axial lengths. Although the two lasers operate at nominally identical sweep rates, a residual up to tens of milliseconds mismatch still exists. To compensate, our custom acquisition software synchronizes the two channels at the volume level, ensuring temporal alignment of anterior and posterior data.

The depth resolution of each channel was characterized by imaging a mirror. The anterior channel achieved a depth resolution of 12.4 µm in air, while the retina channel achieved 8.9 µm in air. The light from the two channels is combined using a custom dichroic mirror (Avantier, North Plainfield, NJ) placed before the final objective lens.

The system also incorporates two NI DAQ boards (PCIe-6361), each driving an independent galvanometer scanner pair (Saturn 5, ScannerMax, Sanford, FL) for its respective channel. Scanner positions are recorded simultaneously at each A-scan, which is essential for accurate 3D reconstruction and distortion correction, as discussed in detail in Section \ref{systemcalibration}.

To facilitate subject alignment, the system integrates a near-infrared pupil alignment camera (acA2050-90um, Basler, Ahrensburg, Germany) (Fig.\ref{fig:optical}(e)) and a custom 840 nm LED illumination ring mounted around the aperture of the final objective lens. A fixation display is also introduced to present fixation targets at eccentricities up to ±12°, allowing data acquisition across a range of gaze angles (Fig.\ref{fig:optical}(f)). The fixation display and alignment camera are combined with the OCT imaging channels using two additional custom designed dichroic mirrors (Avantier, North Plainfield, NJ). The entire optical assembly is mounted on a motorized XY translation stage (Zaber Technologies, Vancouver, BC), and the chin rest is equipped with a motorized Z-axis actuator, enabling precise 3D alignment with the subject's eye (Fig.\ref{fig:system}(b)). 

\subsection{Sample arm design \& performance}
\label{sec:sample_arm}
We designed custom optics for both the anterior and retina channels to meet the specific requirements of anterior and posterior eye imaging. The complete system maintains a working distance of approximately 5 cm, providing adequate clearance for comfortable subject positioning during imaging.

\begin{figure}[h]
    \centering
    \includegraphics[scale =0.9]{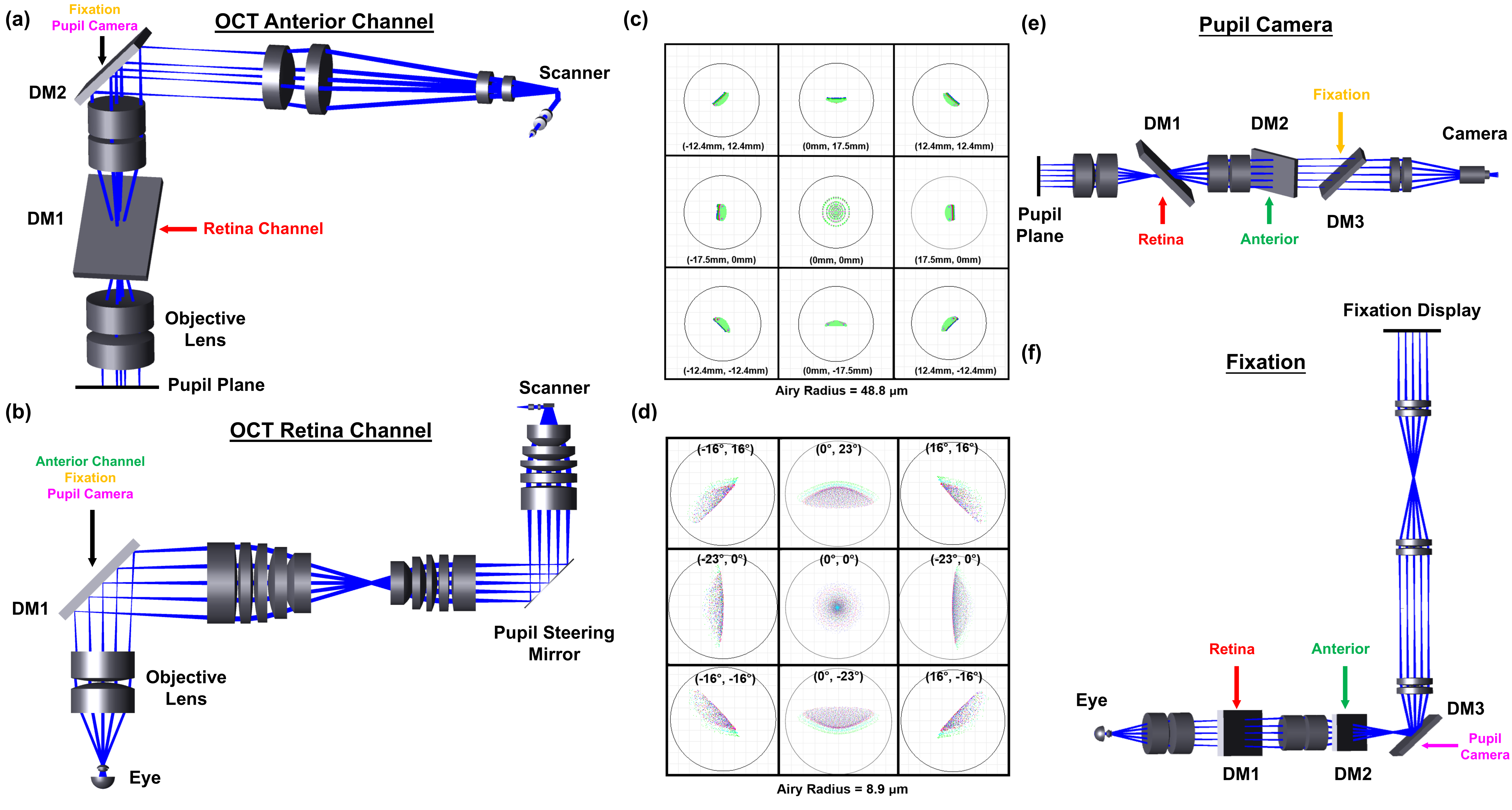}
    \caption{\textbf{Spectrally-multiplexed whole-eye OCT optical design.} (a)OCT anterior channel and (c)corresponding spot diagrams, (b)OCT retina channel and (d)corresponding spot diagrams, (e)pupil alignment camera channel, and (f)fixation display channel }
    \label{fig:optical}
\end{figure}

\textbf{Anterior Channel}: The 1310 nm anterior channel (Fig.\ref{fig:optical}(a)) is designed to provide a wide lateral field of view (FOV) of 35 mm to capture the entire cornea, iris, and sclera. The optical design achieves nearly telecentric scanning and delivers diffraction-limited performance across the entire FOV (Fig.\ref{fig:optical}(c)). The spot size, 48.8 µm, was deliberately chosen to provide a sufficiently large depth of focus to cover the entire anterior eye depth—from the cornea to the crystalline lens, spanning at least 4 mm. The illumination power at the pupil plane is set at 6.7 mW, yielding a peak sensitivity of 107.2 dB, measured on a mirror target.

\textbf{Retina Channel}: The 1060 nm retina channel (Fig.\ref{fig:optical}(b)) is designed to have a lateral FOV of 45° with a diffraction-limited lateral resolution of 8.9 µm using a model eye (Fig.\ref{fig:optical}(d)). The illumination power at the pupil plane is 1.8 mW, resulting in a peak sensitivity of 102.5 dB. To accommodate different refractive errors among subjects, the retina channel incorporates an electrically tunable lens (Optotune, Dietikon, Switzerland) that provides a diopter correction range from -10 D to +8 D. Additionally, an extra 4f relay is incorporated into the retina channel to image the system pupil onto a fast steering mirror (Optics in Motion, Long Beach, CA), enabling pupil steering that can potentially support retina imaging over a maximum gaze range of ±15°. This pupil steering function was not actively used for the current data collection. 

\subsection{Processing pipeline}
We developed a comprehensive, automated processing pipeline to extract accurate whole-eye biometry from the raw OCT volumes. Consistent with prior OCT biometry work \cite{ortiz2010optical,ortiz2011corneal,ortiz2012quantitative,ortiz2013full}, the pipeline consists of three main stages: surface segmentation, distortion correction, surface fitting and refraction correction.

\subsubsection{Surface segmentation and fovea detection}
\label{sec:segmentation}
We developed a deep learning-based pipeline to automatically segment key anatomical boundaries in anterior segment. Prior to training, individual OCT B-scans were preprocessed via log compression and intensity normalization. Trained annotators labeled B-scans using an image annotation platform, with assistance from the Segment Anything Model \cite{ravi2024sam2}. Because the boundary between cornea and sclera is poorly defined in OCT B-scans, they were labeled as a single class during annotation and training. The cornea and sclera are subsequently separated in the processing pipeline using the iris boundary inferred from the iris annotation. All annotations are produced as masks; for downstream refraction correction, these masks are converted to single-pixel surface boundaries—the cornea mask yields the anterior and posterior corneal surfaces, and the iris mask yields the anterior and posterior iris surfaces.

The segmentation model used a DenseUNet—an encoder–decoder network combining U-Net skip connections with DenseNet-style dense connectivity \cite{ronneberger2015u}. The encoder consists of five dense blocks, each containing three convolutional layers (one 3×3 conv followed by two 1×1 bottleneck + 3×3 conv pairs) with dense concatenation, Leaky ReLU activations, and batch normalization at the block output, with 2×2 average pooling between blocks. The decoder uses four upsampling blocks, each with two 1×1 bottleneck + 3×3 conv pairs and skip connections from the encoder. All layers use 32 channels, weights are He-initialized, and a final 1×1 convolution produces the per-pixel output.
 
 We also used a coarse-to-fine two-stage pipeline. Stage 1 localizes the anterior eye region from a low-resolution crop, and Stage 2 refines boundaries on a higher-resolution crop centered on the corneo-scleral region. During training, the crop is centered on the ground-truth annotation; At inference, Stage 1's prediction guides the crop. Both stages use identical DenseUNets trained with weighted cross-entropy loss and AdamW. Training was performed on 51,398 annotated B-scans from 552 OCT volumes (30 subjects), with 7,951 B-scans from 79 volumes (8 subjects) held out for validation.

During model training, data augmentation included brightness, gamma, and contrast jitter, random horizontal flipping, perspective warping, rotation (up to ±20°), and Gaussian noise. Models were optimized with AdamW and selected based on the best Jaccard index (IoU) for segmentation. The final selected model was then exported for deployment in the processing pipeline.

\begin{figure}[h]
    \centering
    \includegraphics[scale =0.6]{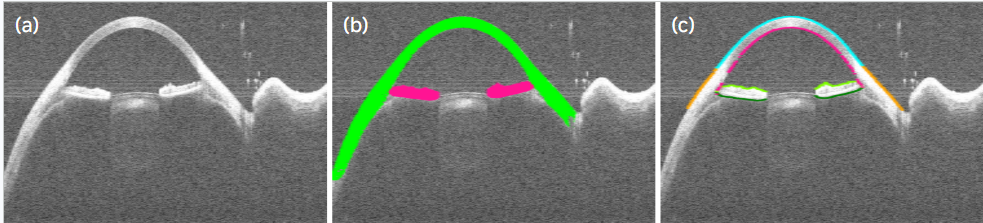}
    \caption{\textbf{Representative anterior OCT segmentation results.} (a)  B-scan input to model and (b) region mask outputs from model. Cornea/sclera in green, iris in magenta.(c) Mask borders, showing split between front surface of cornea (blue) and sclera (orange)}
    \label{fig:segmentation}
\end{figure}

A representative anterior segment OCT B-scan is shown in Fig. \ref{fig:segmentation}(a), along with the segmentation masks produced by our trained model (Fig. \ref{fig:segmentation}(b)). The corneal, iris, and scleral surfaces (Fig. \ref{fig:segmentation}(c)) were subsequently extracted from the mask boundaries and used as inputs to the downstream processing pipeline.

\subsubsection{System calibration}
\label{systemcalibration}
To correct system distortion, we developed a rigorous whole-eye OCT calibration pipeline that establishes a mapping from galvanometer mirror scan voltages to physical 3D coordinates. The calibration process involves imaging a custom dot-pattern target at multiple known depths The calibration target was designed with 2 different dot radii, with a pseudo-random distribution that enables unambiguous identification of lateral position from a single image. A hexapod robot (HXP100-MECA, Newport, Irvine, CA) translates the target to seven distinct Z-positions, and at each position, the OCT system captures a full volume scan. The tunable lens in the retina channel was adjusted to bring the calibration target into focus, unlike during human retinal imaging, where the retina channel operates with a collimated beam into the eye.

\begin{figure}[h]
    \centering
    \includegraphics[scale =0.4]{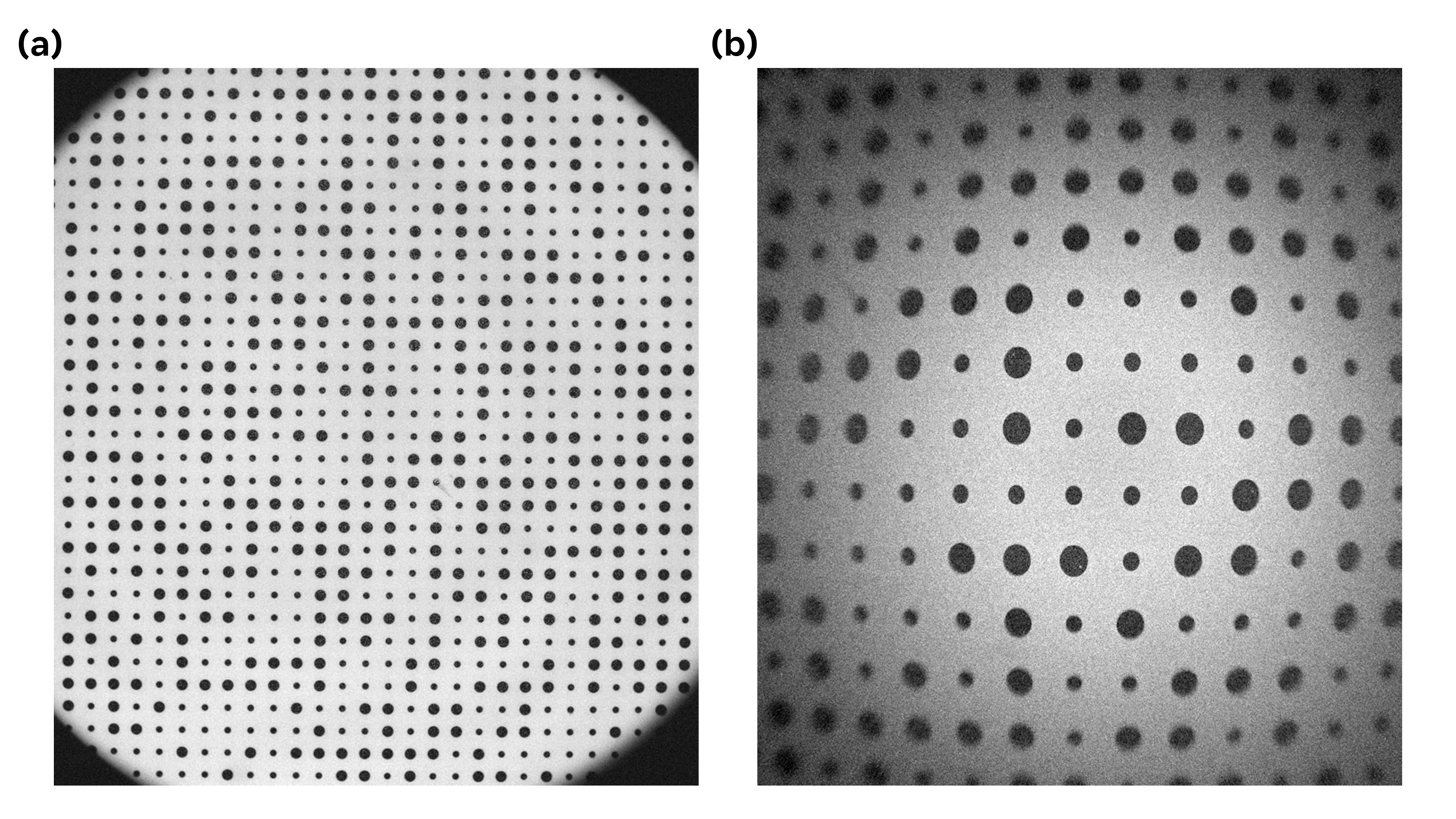}
    \caption{Representative (a) anterior and (b) retina channel en face projection view of custom dot-pattern target}
    \label{fig:calibration}
\end{figure}

The distortion calibration processing pipeline first reconstructs each volume, generates an en face projection image (Fig.\ref{fig:calibration}), and runs edge detection to locate the target surface depth in every A-scan. A dot detection algorithm then identifies the centroids of the known pattern in each en face image, establishing correspondences between pixel coordinates and physical (x, y) positions on the target. The hexapod's commanded Z-offset is applied to obtain the full 3D positions. The core of the calibration is a ray localization algorithm, which leverages the fact that a given pixel observes the target at different depths across the hexapod positions. By collecting the corresponding 3D points for each pixel and fitting a least-squares line through them, the algorithm recovers the origin and direction of the OCT beam ray at that specific scan position.

These per-pixel ray parameters are then expressed as functions of the galvanometer fast- and slow-axis voltages via 3rd-order 2D polynomial fits. This produces a compact mapping function that converts any galvo voltage pair to a full 3D ray. An optional nonlinear refinement step jointly optimizes all polynomial coefficients and the axial pixel size by minimizing the reprojection error across all observations.

Joint channel calibration into a common coordinate system was achieved by translating the calibration target with the hexapod between both channels at different target orientations. Per-channel calibration recovered target position and orientation in the respective local coordinate system, and the relationship between both coordinate systems was computed as a rigid transform. 

\subsubsection{Surface fitting \& refraction correction}
\label{sec:refraction_correction}
Because OCT measures optical path length rather than geometric distance, the apparent positions of structures posterior to the cornea are distorted by refraction at each optical interface. To recover true 3D coordinates, we apply a ray-tracing-based refraction correction to the distortion-corrected point cloud.

Each anatomical interface is modeled with a parametric surface fit to the segmented point cloud. The anterior and posterior cornea surfaces are represented as biconic surfaces. Surface fitting employs RANSAC to reject outliers introduced by segmentation noise, followed by nonlinear least-squares refinement to maximize accuracy.

The pipeline assumes fixed population-average refractive index values rather than per-subject measurements. The indices used are n = 1.376 for the cornea (applied at the air–cornea interface),
and n = 1.336 for the aqueous humor (applied at the cornea–aqueous interface). The cornea is treated as a single homogeneous medium throughout its thickness, rather than modeling its layered microstructure. Snell's law was applied at each interface, and total internal reflection, when encountered, is handled by deactivating the affected ray. The output of this processing step is refraction-corrected anterior point cloud, which serves as the basis for all subsequent biometric measurements, such as cornea topography and 3D pupil center.

\subsection{System validation}
To characterize and validate the performance of our system and processing pipeline, especially system calibration and refraction correction, we designed and performed several studies on precision reference sphere and model eye. 

\subsubsection{System calibration validation using reference sphere}
To investigate the performance limits of OCT in estimating corneal radius and center, we first imaged a 12.5mm radius reference ceramic sphere (RS-DK25-C, Zeiss) that was mounted on the hexapod. The sphere was translated to multiple positions spanning the lateral FOV of anterior OCT channel to evaluate performance across the imaging volume. Intensity-based thresholding was applied to segment the sphere surface from the OCT volumetric data, and a least-squares sphere fitting algorithm was used to estimate the radius and 3D center position. 12 measurements were acquired at 4 lateral positions with 3 repeated acquisitions per position. All measurements used the same scan parameters employed in human eye data collection to ensure consistency and direct comparability of error bounds. Accuracy of radius estimation was evaluated by comparing the OCT-measured radius to the known ground truth value of 12.5 mm. Accuracy of 3D center estimation was assessed by comparing the OCT-measured center displacement between positions against the commanded hexapod translation, which serves as the ground truth reference with sub-micrometer positioning accuracy. Precision was evaluated from the standard deviation of repeated measurements at each position. Performance metrics are reported as median (p50) and 95th percentile (p95) absolute errors.

Overall, our system achieved a median radius error of 14 µm and p95 error of 19 µm, corresponding to relative errors of 0.11\% and 0.15\%, respectively. For 3D center localization, the systems achieved median and p95 errors of around 30 µm. These results demonstrate that our automated processing pipeline, specifically our calibration pipeline, can achieve corneal biometry accuracy at tens of microns level as a ground truth eye biometry measurement platform.

\begin{table}[ht] \centering \begin{tabular}{l c c} \toprule \textbf{Metric} & & \textbf{Absolute error $\mu$m (\%) } \\ \midrule \textbf{Sphere Radius} & p50 & 14 $\mu$m (0.11\%) \\ & p95 & 19 $\mu$m (0.15\%) \\ \addlinespace \textbf{Sphere 3D Center} & p50 & 30 $\mu$m (0.24\%) \\ & p95 & 33 $\mu$m (0.26\%) \\ \bottomrule \end{tabular} \end{table}

\subsubsection{End-to-end pipeline validation using phantom eye}
\begin{figure}[h]
    \centering
    \includegraphics[scale =0.6]{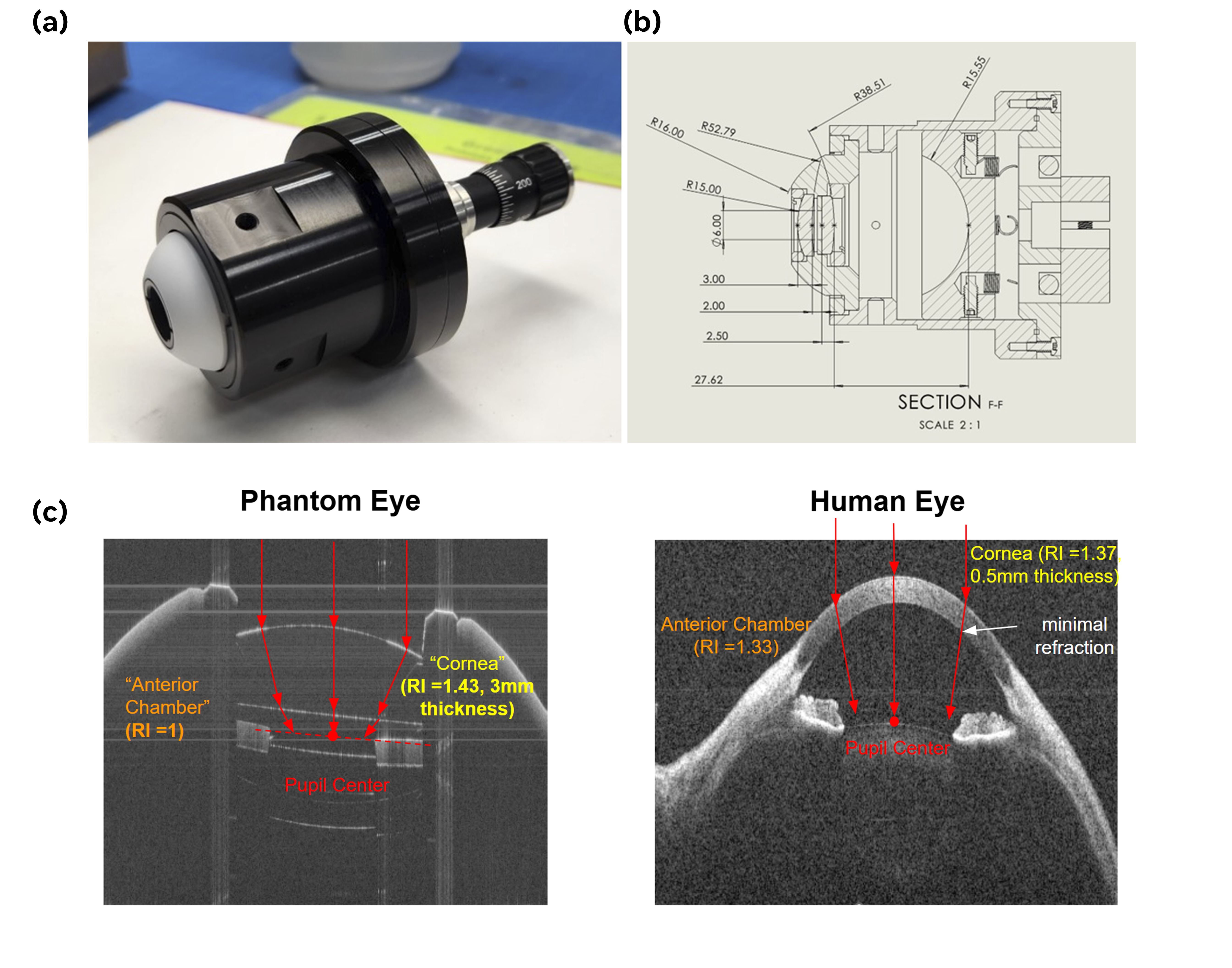}
    \caption{\textbf{Phantom eye study to validate end-to-end pipeline.} (a)Photo and (b)metrology data of the custom high-precision phantom eye, (c) OCT B-scan images of phantom eye vs. human eye}
    \label{fig:phantom eye}
\end{figure}

To evaluate end-to-end performance of our processing pipeline, we imaged a custom high-precision phantom eye mounted on a hexapod positioning stage. The phantom eye (Fig.\ref{fig:phantom eye}(a)) was designed with known geometry and optical properties, featuring a phantom cornea (Thorlabs LE5243, refractive index n = 1.43, central thickness = 3 mm) and a well-defined circular pupil aperture, enabling traceable ground truth measurements (Fig.\ref{fig:phantom eye}(b)).

The phantom eye was imaged under conditions spanning diffrent eye positions and orientations including a combination of yaw and pitch rotations of ±5° and lateral translations of ±5 mm, and a total of 45 OCT volumes were acquired. The OCT processing pipeline was applied to reconstruct the 3D point cloud of the pupil boundary. The pupil center was estimated by fitting an ellipse to the segmented pupil edge and computing its centroid in 3D. Accuracy was evaluated by comparing the OCT-estimated 3D pupil center displacement against the commanded hexapod motion, and reported as median (p50) and 95th percentile (p95) absolute errors.

Overall, the system achieved high accuracy in 3D pupil center estimation across the full range of tested positions and orientations. The system achieved a median 3D pupil center error of 22 µm and a p95 error of 65 µm. No discernible trend of increasing error was observed with added rotation within the tested range of ±5° yaw and ±5° pitch, indicating that the refraction correction algorithm remains robust across various gaze angles.

\begin{table}[ht] \centering \begin{tabular}{l c c c c c} \toprule \textbf{Metric} & \textbf{Mean} & \textbf{STD} & \textbf{p50} & \textbf{p75} & \textbf{p95} \\ \midrule 3D pupil center absolute distance error & 28 $\mu$m & 20 $\mu$m & 22 $\mu$m & 42 $\mu$m & 65 $\mu$m \\ \bottomrule \end{tabular} \end{table}

These results are expected to represent a conservative (upper-bound) estimate of performance in human eyes for two reasons (Fig.\ref{fig:phantom eye}(c)): (1) the phantom eye exhibits a substantially larger refractive index difference between the cornea and anterior chamber ($\Delta$ n $\approx$ 0.43) compared to the human eye ($\Delta$ n $\approx$ 0.04), resulting in more severe refraction requiring correction; and (2) the phantom cornea is approximately 5× thicker (3 mm) than the human cornea (500–600 µm), further amplifying refraction-induced geometric distortions. The successful correction of these exaggerated optical effects demonstrates the robustness of the refraction correction pipeline for in-vivo measurements.

\subsection{Data Collection}
\subsubsection{OCT imaging parameters}
We used custom-developed OCT data acquisition software and implemented a triangular raster scanning waveform for both channels, acquiring 500 A-scans per B-scan and 250 B-scans per volume, corresponding to a volumetric rate of approximately 1.6 Hz. The anterior segment channel has a field of view (FOV) of approximately 35 mm × 17.5 mm, with nearly uniform horizontal and vertical sampling density. The retinal channel has a FOV of approximately 36° × 18°, also with nearly uniform sampling density in both directions

\subsubsection{Data-collection protocol}
\label{sec:data_collection}
Participants were given a consent form approved through internal ethics review and a demographic survey. Screening criteria and study demographics are described in Section \ref{sec:demographics}.

Participants were seated in front of the scanner resting their head on a rigid chinrest with a forehead bar, and the study facilitator aligned the chinrest with a motorized stage to the scan position. Optimal eye position was determined using an integrated pupil camera which shares the same axis as the OCT scanner, described in detail in Section \ref{sec:oct_engine}. Once the eye position was aligned, the study facilitator then tuned the diopter correction (described in Section \ref{sec:sample_arm}) of the retina channel scan to achieve OCT retinal imaging with best focus. This was done using the interactive GUI which shows a live preview of the B-scans captured by the system. 

Lighting in the lab began at nominal indoor lighting (100 lux) and during the experiment was adjusted to darkness (0 lux) and to a bright lighting condition (1000 lux) measured on a portable hand-held light meter at the approximate location of the participant’s eye in the chinrest. Lighting adjustments were made to prompt pupil dilation and contraction. After the study facilitator observed that the pupil dynamics had stabilized post-light adjustment, realignment was performed as needed and captures proceeded. 
 
 \begin{table}[ht] \centering \caption{Experimental Conditions Matrix} \label{tab:conditions} \begin{tabular}{llccc} \toprule \textbf{Lighting} & \textbf{Eye Condition} & \textbf{Target} & \textbf{Positions} & \textbf{Volumes} \\ & & & & \textbf{(per position)} \\ \midrule \multirow{3}{*}{Bright} & \multirow{2}{*}{Wide Eyes} & Central & 1 & $3 \times 2 = 6$ \\ & & Peripheral & 8 & $3 \times 2 = 6$ \\ \cmidrule(l){2-5} & Normal Eyes & Central & 1 & $3 \times 2 = 6$ \\ \midrule \multirow{2}{*}{Standard} & Wide Eyes & Central & 1 & $3 \times 2 = 6$ \\ \cmidrule(l){2-5} & Normal Eyes & Central & 1 & $3 \times 2 = 6$ \\ \midrule \multirow{2}{*}{Dark} & Wide Eyes & Central & 1 & $3 \times 2 = 6$ \\ \cmidrule(l){2-5} & Normal Eyes & Central & 1 & $3 \times 2 = 6$ \\ \midrule \multicolumn{3}{l}{\textbf{Total unique conditions:}} & \multicolumn{2}{c}{\textbf{7}} \\ \multicolumn{3}{l}{\textbf{Total volumes per participant:}} & \multicolumn{2}{c}{$\mathbf{84}$} \\ \bottomrule \end{tabular} \vspace{6pt} \footnotesize 
 \\
  \textit{Note:} Peripheral targets comprise 8 gaze directions (L, UL, U, UR, R, LR, D, LL). Each position is repeated 3 times with 2 volumes per repeat. Peripheral fixations were collected only for the Bright/Wide Eyes condition. \end{table}

Participants were instructed that during the task they should look at the target and keep their eyes still, avoiding blinking or moving. In some of the trials, they were instructed to open their eyes wide as if they had been surprised. The full matrix of experimental conditions is in Table \ref{tab:conditions}. The experiment facilitators observed eye stability in the pupil camera view and would repeat sessions as needed in the case of excessive movement or blinking. Simultaneous scans of the right eye retina and anterior segment were collected at central fixation, and at 8 points around a 9 degree ring. At each point, participants were asked to repeat their fixation three times, blinking between captures. Capture was initiated by the facilitator after observing the participant’s eye was still. Each capture included two repeated volumes to further minimize the risk of motion during a scan, leading to a total of six volumes at each fixation point.

\subsubsection{Data quality assurance pipeline}
\label{sec:QA}
\begin{figure}[h]
    \centering
    \includegraphics[width=0.5\linewidth]{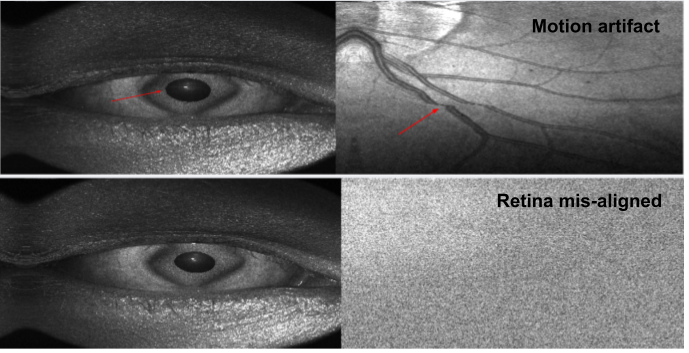}
    \caption{\textbf{Example cases that would be rejected by the quality assurance pipeline.} Data was shown to annotators as pairs of en-face projections from the anterior and retina scans. Top: example showing motion artifact marked by red arrows. Bottom: example of mis-aligned retina channel, where retina is not in focus}
    \label{fig:qa_example}
\end{figure}

Volumes were manually inspected as part of a quality assurance (QA) process. Trained graders inspected the en-face view of the retina and rejected volumes where the retina was obscured or showed motion artifacts. 33\% of the data were rejected due to participant motion, and 8\% of the scans were rejected due to poor retina visibility. Some representative cases that were rejected by the QA pipeline are shown in Fig.\ref{fig:qa_example}. Overall, 45\% of the volumes collected were acceptable, with retina visibility rejections stemming primarily from the downward gaze directions and from the bright lighting condition where the pupil was small and more difficult to align.

\section{Data description}

\subsection{Demographic breakdown}
\label{sec:demographics}

Participants were recruited from the Redmond, WA area and compensated for their participation in the study. Screening criteria included a question about the participants' vision prescription, and participants who had Rx stronger than $\pm$ 5.00 diopters and did not wear contact lenses were excluded from the study. This was done to ensure users could properly see gaze targets used for fixation during a follow up portion of the study not included in this dataset. For participants who did wear contact lenses, scans were conducted with and without the lenses in place. Of the final cohort who passed quality assurance checks, 55\% were female.

\begin{figure}[htbp] \centering \begin{minipage}{0.5\textwidth} \centering \includegraphics[width=\linewidth]{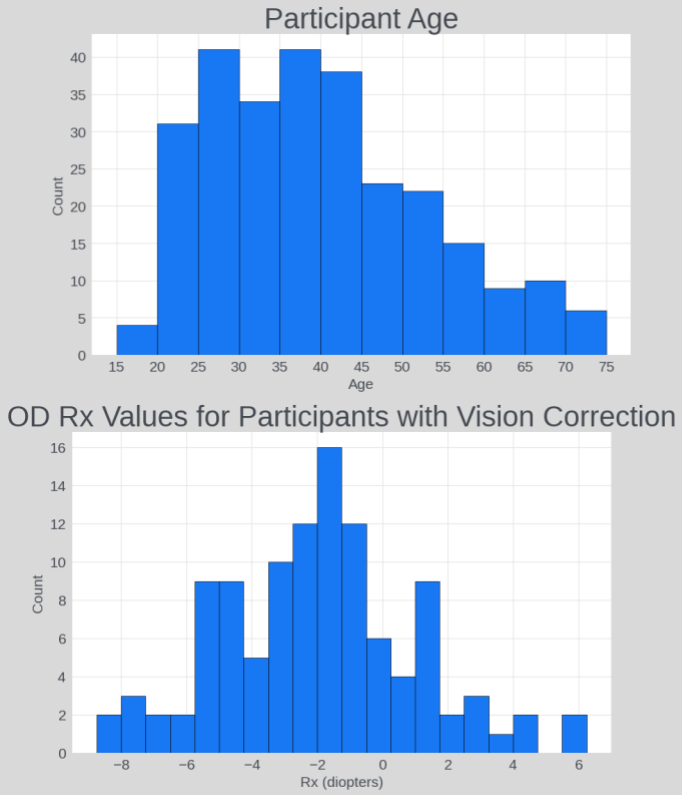} \end{minipage} \hfill \begin{minipage}{0.45\textwidth} \centering \small \setlength{\tabcolsep}{3pt} \begin{tabular}{lrr} \toprule Ethnicity & Count & Percent \\ \midrule White & 130 & 47.45 \\ East Asian & 43 & 15.69 \\ South Asian & 25 & 9.12 \\ Southeast Asian & 25 & 9.12 \\ Hispanic/Latino & 19 & 6.93 \\ Black/African American & 11 & 4.01 \\ Middle Eastern & 10 & 3.65 \\ Other & 5 & 1.82 \\ Native Hawaiian/Pacific Islander & 4 & 1.46 \\ American Indian/Alaska Native & 2 & 0.73 \\ \bottomrule \end{tabular} \label{fig:demographics} \end{minipage}\caption{Demographics summary. Participants had mean age 39.5 $\pm$ 13.3 years. 44\% (122 participants) in the dataset reported having vision correction. The distribution of their prescription strength is shown here in the lower panel}  \end{figure}

\subsection{Representative imaging results \& 3D anterior point clouds}

\begin{figure}[h]
    \centering  
    \includegraphics[scale =0.4]{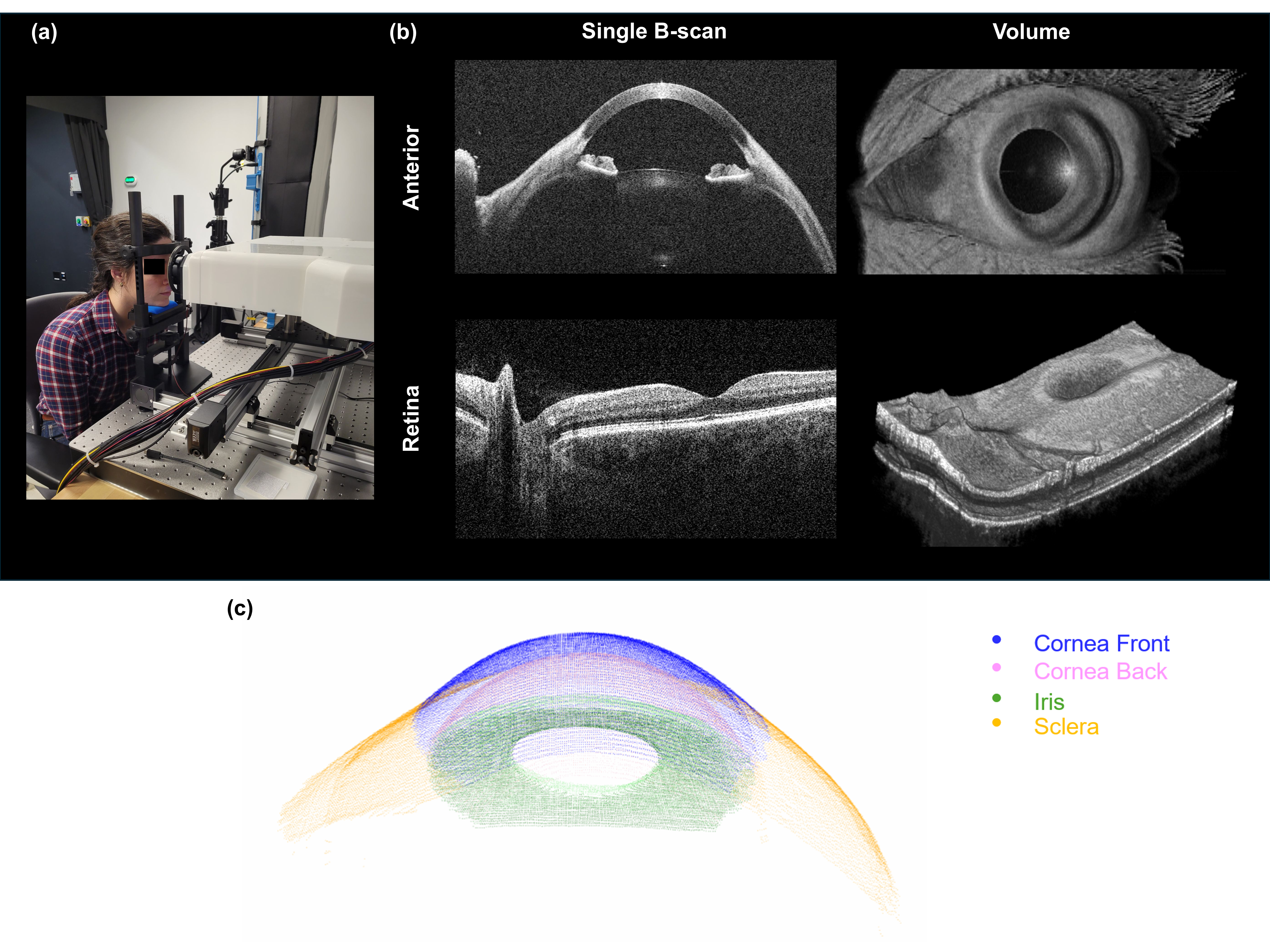}
    \caption{\textbf{Representative whole-eye OCT human imaging results.} (a)Photo of a participant aligned with the system, (b) representative single B-scans and volume renderings of both imaging channels and (c) representative processed 3D anterior point.}
    \label{fig:imagingresults}
\end{figure}

Figure \ref{fig:imagingresults} presents representative imaging results and processed point clouds from our data collection. Figure \ref{fig:imagingresults}(a) shows a participant aligned with the whole-eye OCT system. Figure \ref{fig:imagingresults}(b) displays representative single B-scans and volumetric renderings from both anterior segment and retinal channels. Figure \ref{fig:imagingresults}(c) illustrates the 3D anterior point cloud generated by our end-to-end processing pipeline

\subsection{Open-source dataset}
\label{sec:dataset}
A dataset of 6,621 raw and processed volumes from 276 unique participants can be accessed at \url{https://github.com/facebookresearch/rl_whole_eye_oct}. For each participant, at least 5 repeated volumes have been captured while the user fixated at a central target. These data have passed manual QA as described in Section \ref{sec:QA}. In the dataset there are three categories of data: raw volumes, processed data, and manual annotations. The raw volumes are the raw spectral OCT B-scan stacks paired with the scan waveform used during capture. The processed data category contains processed B-scans, segmentation masks for each B-scan and the boundary pixels for each mask. The processed data category also contains the full calibrated point cloud with segmentation labels for each volume as shown in Figure \ref{fig:imagingresults}(c). The manual annotations category contains human-labeled segmentation masks that were used to train our automatic segmentation model described in Section \ref{sec:segmentation}. This category contains a subset of data from 34 participants, and annotations of each B-scan are provided.

\section{Acknowledgement}
We would like to thank Christian Viehland and Hafeez Al-Dhalla from Coherence Consulting for their integral contribution in the design and development of our whole-eye OCT system.

\vspace{1cm}
\noindent *These authors contributed equally to this work

\bibliographystyle{ieeetr}
\bibliography{references}

\end{document}